\documentclass{jfm}
\usepackage{graphicx}
\usepackage{natbib}
\usepackage{psfrag} 
\usepackage[latin1]{inputenc} 
\usepackage{amsmath}
\usepackage{bm}
\ifCUPmtlplainloaded \else
   \checkfont{eurm10}
   \iffontfound
     \IfFileExists{upmath.sty}
       {\typeout{^^JFound AMS Euler Roman fonts on the system,
                    using the 'upmath' package.^^J}%
        \usepackage{upmath}}
       {\typeout{^^JFound AMS Euler Roman fonts on the system, but you
                    dont seem to have the}%
        \typeout{'upmath' package installed. JFM.cls can take advantage
                  of these fonts,^^Jif you use 'upmath' package.^^J}%
       }
   \else
   \fi
\fi

\ifCUPmtlplainloaded \else
   \checkfont{msam10}
   \iffontfound
     \IfFileExists{amssymb.sty}
       {\typeout{^^JFound AMS Symbol fonts on the system, using the
                 'amssymb' package.^^J}%
        \usepackage{amssymb}%
          
        \let\ge=\geqslant  
       }{}
   \fi
\fi

\ifCUPmtlplainloaded \else
   \IfFileExists{amsbsy.sty}
     {\typeout{^^JFound the 'amsbsy' package on the system, using it.^^J}%
      \usepackage{amsbsy}}
     {}
\fi

\newcommand{\eq}{\begin{equation}}
\newcommand{\qe}{\end{equation}} 

\title{Forced dewetting on porous media}

\author[Olivier Devauchelle, Christophe Josserand and Stéphane Zaleski]%
{OLIVIER DEVAUCHELLE,
CHRISTOPHE JOSSERAND
and STEPHANE ZALESKI
}
\affiliation{Laboratoire de Modélisation en Mécanique, 
CNRS-UMR 7606, Case 162, 4 place Jussieu, 75252 
Paris C\'edex 05-France\\}

\pubyear{2005}
\volume{xxx}
\pagerange{xxx--yyy}
\begin{document}
\date \today
\maketitle

\begin{abstract}

We study the dewetting of a porous plate withdrawn from a bath of fluid. The microscopic contact angle 
is fixed to zero and the flow is assumed to be parallel to the plate (lubrication approximation). The 
ordinary differential equation involving the position of the water surface is analysed in phase 
space by means of numerical integration. We show the existence of a critical value of the capillary number 
$\eta U / \gamma$, above which no stationary contact line can exist. An analytical model, based on asymptotic 
matching is developed, that reproduces the dependence of the critical capillary number on the angle of the plate with respect to the horizontal for large control parameters ($3/2$ power law).

\end{abstract}

\section{Introduction}
When sea retreats from the shore, sand structures appear as solid granular
particles are transported \emph{via} the liquid. Liquid motion and particularly film 
retraction on an erodible medium are known to create impressive erosion patterns, such 
as sand ripples for oscillatory waves \cite[]{stegner,scherer} or sand furrows \cite[]{dae03,roth04}.
The case of liquid retraction from a granular bed can be understood as a dewetting dynamics on a porous
erodible bed. Such physical phenomena have been reproduced in the laboratory by pulling a plate covered with a bed of grains out of a liquid tank \cite[]{dae03}. This situation is similar to the well-known experiment investigating a moving contact line on a non-porous plate \cite[]{Blake79}. In this latter case, a
contact line exists for small removal speed $U$, whereas for higher speed (above a well-defined critical value
$U_{cr}$) a macroscopic water film (the so-called Landau-Levich-Derjaguin film (denoted LLD later on), see \cite{lan42,Derja43}) 
covers the whole plane \cite[]{egg04}. We propose here to
investigate this transition for a saturated porous medium, in connection with recent experiments
involving granular materials by \cite{dae03}. There, a
motor-driven plane, covered with a granular layer, is withdrawn from a water tank
at constant speed $U$. The solid plane is tilted to an angle $\theta$. At high enough velocity, erosion river 
networks and mudflows are observed, whereas only light patterns appear at smaller speed. We investigate
the loss of a static contact line and seek to relate it to the transition between various erosion regimes. We 
therefore seek the critical velocity above which no 
static contact line can exist on a granular bed. Below this critical velocity, almost no grain motion is 
observed so that we identify the granular bed with a rigid porous medium.
Dewetting on a porous medium has already been studied in different configurations (see \cite{rap99,arad00,babr01} 
and references herein).
The case of a porous plate removed from a liquid was studied by \cite{rap99}, but
the focus was on the spatial evolution of the LLD film, the existence of which was assumed.
The contact line dynamics was also studied \cite[]{arad00} for a wet horizontal support in which the liquid
is sucked in the dry porous medium.

From a more theoretical point of view, the problem of a moving contact line on a
porous solid is pointed out by \cite{gen85} as a natural
regularization for the contact line dynamics equations. The bulk
liquid flow through the porous solid indeed removes the usual stress singularity
that one would encounter at a contact line with a no-slip condition \cite[]{duss74}. No additional assumption, such as the introduction of a Navier slip at microscopic scale, is then required. However, another question arises when considering a porous medium: what is the relevant condition for the contact angle at the contact line? As discussed below, we propose here for a saturated system to take a zero contact angle. 

The paper is organized as follows: in the next Section, we use the lubrication approximation to deduce the equation for the static interface shape, both for the contact line and a zero-flux LLD film. In Section \ref{numerical}, we exhibit the transition between these 
two configurations as the pulling velocity increases by means of a shooting method. Then, we propose to 
interpret the solutions in the framework of dynamical systems (Section \ref{dynamical}).

\begin{figure}
\begin{center}
\psfrag{theta}{$\theta$}
\psfrag{x}{$x$}
\psfrag{y}{$y$}
\psfrag{h(x)}{$h(x)$}
\psfrag{U}{$\bm{U}$}
\psfrag{g}{$\bm{g}$}
\includegraphics[width=9cm]{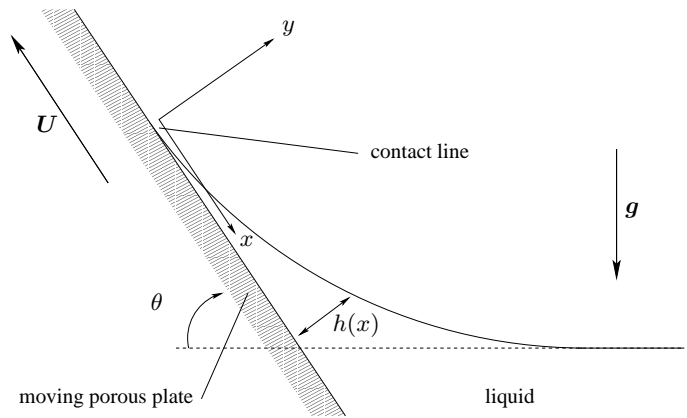}
\caption{A porous plate of conductivity $k$ is being withdrawn from a
liquid bath with speed $\bf{U}$ at angle $\theta$.}
\label{schema}
\end{center}
\end{figure}

\section{Principles}
\subsection{Lubrication approximation}

Our approach seeks to determine the velocity (if any) above which the static contact line can no longer exist in a granular bed withdrawal experiment. Below this velocity, we can consider that the grains almost do not move relatively to the withdrawn plate. Thus the granular material is represented by a non-erodible porous medium (see
Figure \ref{schema}) of permeability $k$, and we only have to investigate the stationary problem. The fluid is characterized by its density $\rho$, dynamical viscosity $\eta$ and surface tension $\gamma$. 
Assuming invariance in the $z$-direction, we consider the two-dimensional problem where the water surface is described by the function $h(x)$. For the plate velocities pertinent 
to the problem (typically $0.5 \: \text{cm} . \text{s}^{-1} $) and the estimated porosity of the granular bed ($k \approx 10^{-12} \: \text{m}^2$) 
we can consider that the porous medium remains fully saturated with water at any distance from the free water level.

We will restrict our analysis to small angles $\theta$ so that the lubrication approximation can be employed (\emph{i.e.} $\theta \; , \; \| h' \| \ll 1$ where the $'$ stands for the $x$-derivative). Only the $x$-component $u$ of the velocity has to be taken into account, for which Poiseuille profile is assumed, with a vanishing tangential stress on the gas side:

\eq \left. \frac{\partial u}{\partial y} \right| _{y=h(x)} = 0. \label{nostress} \qe

Another boundary condition has to be written at the porous surface. The classical no-slip condition, as required at the solid-fluid interface on an impermeable plate, leads to the following equation for $h(x)$ (its derivation is similar to the one presented in Appendix \ref{MoveEq}):

\eq h''' - h' + \theta = \frac{3 \text{Ca} }{h^2} \label{noslip}, \qe

where lengths $h$ and $x$ have been made dimensionless by the capillary length $l_c=\sqrt{\gamma / \rho g}$ ($l_c \approx 2.8 \: \text{mm}$ for water) and $\text{Ca} $
is the capillary number, defined by

\[\text{Ca}=\frac{\eta U}{\gamma}.\]

Equation (\ref{noslip}) and any derivative of its solutions are singular at the contact line, where $h=0$ 
\cite[]{duffy}. For a non-porous surface, a short-length regularization is invoked coming either from effective 
slip near the contact line \cite[]{huh71}, the existence of a pre-wetting liquid film and Van der Waals 
forces \cite[]{gennes2,hervet} or a "diffuse interface model" \cite[]{seppecher}. Such a regularization always 
involves a microscopic cut-off length (on the order of $1$ nm) below which it is claimed that hydrodynamics fails. 
The Navier slip condition is then mostly used in numerical simulations investigating moving contact line 
problems \cite[]{renard}. This condition reads at the solid-fluid interface: $ u-U=\Lambda_N \partial u / \partial y$ at $y=0$ where $\Lambda_N$ is the cut-off length. If $\lambda_N$ is the rescaled cut-off length (that is $\lambda_N = \Lambda_N / l_c$), Equation (\ref{noslip}) becomes

\eq h''' - h' + \theta = \frac{\text{Ca} }{h^2/3+\lambda_N h}  \label{slip}. \qe 

Equation (\ref{slip}) can be numerically solved and analytically approached. A contact line is then found to exist as long as the capillary number is smaller than a critical value, above which a macroscopic LLD film is deposited on the solid \cite[]{egg04}. Notice however that not all the singularities discussed above are suppressed by the Navier slip condition since the capillary pressure still diverges at the contact line (see Appendix \ref{DivergenceDemo}).

\subsection{The case of porous solid}

A porous solid allows for both interfacial slip (first proposed by \cite{bea67}) and bulk flow. Using the Brinkman
equation to describe the flow inside the porous medium, \cite{nea74} showed that, for a homogeneous porous media, the magnitude of the slip is proportional to the prevailing shear stress:

\eq \left. u  \right| _{y=0} - \left. u_p \right| _{y=0} = \frac{\sqrt{k}}{\alpha} \left. \frac{\partial u}{\partial y} \right| _{y=0}, \label{slipporous} \qe

where $k$ is the permeability of the solid, $\alpha$ a coefficient of order one, and $u_p$ the velocity of
the fluid in the porous medium. Darcy's law holds in the solid so that

\eq u_p + U =-\frac{k}{\eta} \left( \frac{\partial p}{\partial x} - \rho g \theta \right), \label{darcy}\qe

and finally the Equation hereafter describes the shape of the steady fluid film under withdrawal (the detailed derivation is presented in Appendix \ref{MoveEq}):

\eq h''' - h' + \theta = \frac{\text{Ca} }{h^2/3+ \lambda h / \alpha +\lambda^2} \label{total}, \qe

where $\lambda=\sqrt{k}/l_c$. Such an equation is similar to those studied for the contact line on a solid surface, using specific boundary conditions at the solid surface \cite[]{egg04,hock01}. It has been shown (in the case of a plate \emph{pushed} into water) that the details of the regularization do not influence the far-field fluid flow as long as the cut-off length is small enough \cite[]{egg04b}. However, an important difference in our case lies in the typical values of $\lambda$ involved in porous media ($\approx 10^{-2}$ in \cite{dae03}) to be compared with $10^{-6}$ for regular solids.

Here we would like to point out that in recent papers \cite[]{tab03,hadj03} a second order-slip law is used to model the flow of gases at large Knudsen numbers. This boundary condition (adapted to the present notations) reads

\[ \left. u  \right| _{y=0} + U = 
\Lambda_C \left. \frac{\partial u}{\partial y} \right| _{y=0} -
\alpha_C \Lambda_C \left. \frac{\partial^2 u}{\partial y ^2} \right| _{y=0}, \]

where $\alpha_C$ is a positive coefficient of order one, and $\Lambda_C$ is a slip length of the same order than 
the mean free path of the gas. If such a boundary condition were used in the case of a contact line, again 
Equation (\ref{total}) would be obtained.

\subsection{Boundary conditions}

The limit for large positive values of $x$ is well-defined: the water surface is horizontal far from the plane, that is

\eq h(x) \mathop{\sim}_{x \rightarrow \infty} \theta x  \label{BCip} .\qe

For the two remaining boundary conditions, two different cases will be studied, depending on whether a contact line is formed between the water surface and the solid plate, or if a film of water remains on the solid surface. In the first case, the water level vanishes at the origin and a contact angle $\theta_0$ is usually imposed; the \emph{contact line set} of boundary conditions is

\eq \left\{
\begin{array}{c}
h(0)=0,\\
h'(0)=\theta_0.\\
\end{array} 
\right. \label{contlineset} \qe 

Following \cite{rap99}, we will hereafter consider that the contact angle $\theta_0$ is zero for dewetting on a porous media. We argue indeed that for saturated porous media, the liquid film wets completely the surface, leading to an effective zero contact angle.

When LLD film starts at the meniscus, the only boundary condition known \emph{a priori} is

\eq \lim_{x \rightarrow -\infty} h(x)=h_f \label{nusselt} \qe

where $h_f$ is a constant solution of Equation (\ref{total}).
Notice that Equation (\ref{total}) stands only for zero-flux films, that neither add nor withdraw water from the tank. We omit the important case where a LLD film is continuously growing with time, a case that was investigated by \cite{hock01}. Our study is therefore relevant to determine the loss of a static contact line solution, without any information about the dynamics. The stability of the solution as well as the time-dependent dynamics of a moving meniscus cannot be studied at this stage and will be the purpose of further work. However, we will see in Section \ref{dynamical} that the film solutions of (\ref{total}) satisfying (\ref{nusselt}) play an important role in the dynamical system describing our solutions. Finally, for a solution to be acceptable, the water level must always lay above the porous medium: $\forall x , \; h(x)>0$.

\subsection{Parameters}

The parameter $\alpha$ comes from the detailed modelisation of the interface slip flow \cite[]{bea67,lhui03}, and varies generally between $0.1$ and $4$ \cite[]{nea74}. Notice for instance that if $\alpha$ is smaller than $\sqrt{3}/2$, Equation (\ref{total}) becomes singular for some negative values of $h$. However, for the sake of simplicity the coefficient $\alpha$ is fixed to one in the present study. Now, for $\alpha=1$, Equation (\ref{total}) is an ordinary differential equation (ODE) with three parameters: $\theta$, $\text{Ca}$ and $\lambda$. In fact, one should notice that this equation is only a two-parameter ODE: defining $h_*=h/\lambda$, Equation (\ref{total}) becomes

\eq h_*''' - h_*' + \theta^* = \frac{\text{Ca}^*}{h^{2}_*/3+h_*+1} \label{totalrs} ,\qe 

where $\theta^*$ and $\text{Ca}^*$ are defined as follow:

\[ \theta^* = \theta/\lambda = \theta\sqrt{\frac{\gamma}{\rho g k}}, \]
\[ \text{Ca}^* = \text{Ca}/\lambda^3 = \frac{U \eta \sqrt{\gamma}}{(\rho g k)^{3/2}}.\]

Equation (\ref{totalrs}) is the one we will study later on, but we will omit the $_*$ on $h_*$ for the sake of readability.
These nondimensional parameters were choosen because they are proportional to the two experimental parameters that may be easily and continuously tuned, namely $U$ and $\theta$. If the permeability of the porous solid is extremely low, both $\theta^*$ and $\text{Ca}^*$ tend to infinity, as well as the ratio $\text{Ca}^*/\theta^* = U \eta / (\theta \rho g k)$. In this case, the velocity inside the porous medium is extremely slow compared to $U$. For the experimental study of \cite{dae03}, the rescaled capillary number $\text{Ca}^*$ is of the order of $10^5$.

\subsection{Hydrostatic solutions}\label{hydrosol}

Any solution which respects the boundary condition (\ref{BCip}) for large $x$ verifies

\[ \lim_{x \rightarrow + \infty} h(x) = + \infty ,\]

thus for large $x$, Equation (\ref{totalrs}) becomes

\eq h''' - h' + \theta^* = 0. \label{eqi}\qe 

The behaviour of the water surface at large $x$, hereafter denoted by $h_{\infty}$, is directly obtained from Equation (\ref{eqi}):

\eq h_{\infty}(x) = A_{\infty} + \theta^* x + (\theta^* - \theta^*_{\text{ap}}) \exp{(-x)} \label{inftysol} , \qe

where $A_{\infty}$ and $\theta^*_{\text{ap}}$ are two constants, corresponding respectively to the length of the dynamical meniscus and to the so-called \emph{apparent contact angle} (note that the dimensional apparent contact angle is actually $\theta^*_{\text{ap}} \lambda$). Even though the lubrication approximation is not expected to hold for large $x$ since the water level is not small anymore, Equation (\ref{eqi}) leads to the classical static meniscus solution (remember that $x$ has been scaled by the capillary length $l_c$). Consequently, we may consider that Equation (\ref{totalrs}) holds at any position on the $x$-axis.

\section{Numerical results}
\label{numerical}
\subsection{Contact line solutions}

To seek steady contact line solutions, Equation (\ref{totalrs}) may be solved numerically using a finite-difference algorithm. In the case of a contact
line, two boundary conditions may be fixed at $x=0$ by the contact line conditions (\ref{contlineset}). The third condition comes from the flat water level at infinity (\ref{BCip}). With the new notations, we end up with the following system:

\[h''' - h' = f(h) , \]
\eq h(0)=h'(0)=0 \label{finalset} , \qe
\[h(x) \mathop{\sim}_{x \rightarrow \infty} \theta^* x ,\]

where

\[ f(h)=\frac{\text{Ca}^* }{h^2/3+h+1} - \theta^* . \]

\begin{figure}
\begin{center}

\psfrag{x}{$x$}
\psfrag{h}{$h$}
\psfrag{legA}{$\text{Ca}^*=1.2$}
\psfrag{legB}{$\text{Ca}^*=0.1$}
\psfrag{legC}{$\text{Ca}^*=0.8$}
\psfrag{legD}{$\text{Ca}^*=1.0245$}
\includegraphics[width=10cm]{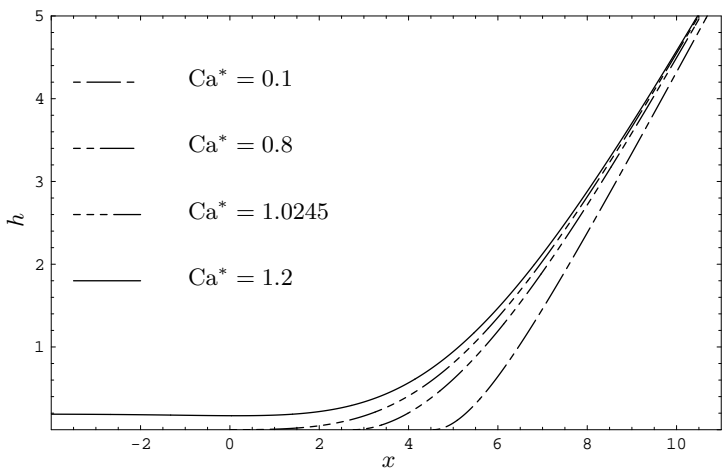}
\caption{Numerical solutions of Equation (\ref{totalrs}) for different capillary numbers. The rescaled tilt angle $\theta^*$ is fixed to $1$}
\label{contlinesol}
\end{center}
\end{figure}

We use a shooting method (see \cite{man90}), varying the initial curvature $h''(0)$ in order to find the numerical solution which corresponds to the hydrostatic condition at large $x$. Some numerical contact line solutions to Equation (\ref{finalset}) are presented on Figure \ref{contlinesol} for different $\text{Ca}^*$ at fixed $\theta^*$.

\begin{figure}
\begin{center}

\psfrag{c}{$h''(0)$}
\psfrag{Ca}{$\text{Ca}^*$}
\psfrag{Cacr}{$\text{Ca}^*_c$}
\includegraphics[width=10cm]{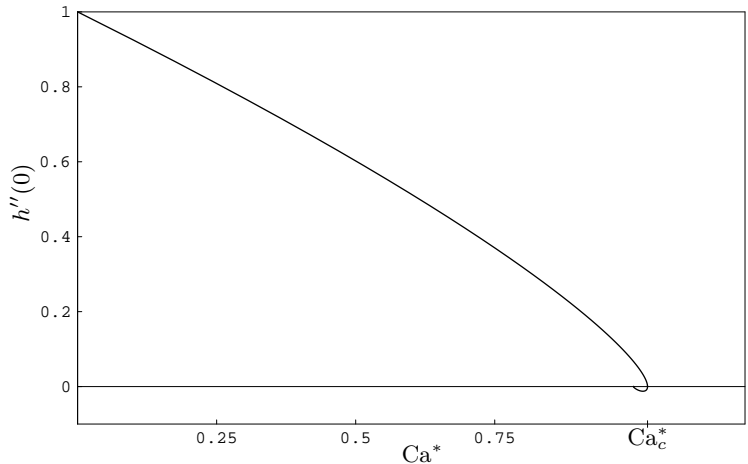}
\caption{Curvature at the origin versus rescaled capillary number, for a contact line solution of Equation \ref{totalrs}. The rescaled tilt angle $\theta^*$ is fixed to $1$. No solution is found for capillary numbers higher than $\text{Ca}^*_c \approx 1.0247$.}
\label{hopf}
\end{center}
\end{figure}

As shown on Figure \ref{contlinesol}, the contact line zone is somehow streched as the capillary number is increased. In other words, the curvature $h''(0)$ at the origin
tends to zero as $\text{Ca}^*$ tends to a critical value $\text{Ca}^*_c$. Above this critical value, no contact line solution can be found by the shooting mehod. This transition will be clarified below using the dynamical system associated to (\ref{finalset}). The disappearance of the contact line solution may be represented in a kind of bifurcation diagram, plotting the curvature at the origin against the capillary number, as shown in Figure \ref{hopf}. Notice that, even though the curvature at the origin tends to zero as $\text{Ca}^*_c$ is approached, the contact line solution does not become unrealistic for $\text{Ca}>\text{Ca}^*_c$ owing to a negative initial curvature, but rather disappears by a bifurcation. Above the critical capillary number, no matching exists between the behavior of the solution at the contact line and the gravity-capillary solution.

\subsection{Film solutions}

A LLD solution can exist whenever there is a positive value $h_f$ such that $f(h_f)=0$, which occurs as soon as $\text{Ca}^* \ge \theta^*$. Two major limitations have to be pointed out for these film solutions: first, we might not be able to match this film solution to the hydrostatic region with $h$ remaining positive everywhere. Moreover we restrict our analysis here to an already-established film of zero mass flux, whereas  transitory and/or finite flux solutions should be considered \cite[]{hock01}, and are likely to exist for smaller capillary numbers. Numerically, above the critical capillary number $\text{Ca}^*_c$, we have always been able to find a zero flux LLD film of 
thickness $h_f$ in the limit $x \rightarrow -\infty$ that could match to the hydrostatic solution without crossing $h=0$ (see Figure \ref{contlinesol}). Such a solution may be numerically approached, using a special shooting method described in Section \ref{speshootmet}. We observed that as the capillary number is decreased, the film surface is shifted down along the $y$-axis (see Figure \ref{ContactAtMinimum}), and we may define a second critical capillary number $\text{Ca}^*_{c,2}$, bellow which the film solution becomes negative in some region. Consequently, if $\text{Ca}^*_{c,2}$ is smaller than $\text{Ca}^*_c$, hysteresis may occurs, that is, two solutions, a contact line one and LLD film one, may co-exist for same tilt angle and capillary number.

\section{Dynamical systems interpretation}
\label{dynamical}

\subsection{Phase space}

In what follow, we interpret and develop the preceding results using dynamical system theory \cite[]{str94}. Let us consider the phase space ${\cal V}$ corresponding to Equation (\ref{totalrs}), that is, $\mathbb{R}^3$ with coordinates $(h,h',h'')$. Any solution of (\ref{totalrs}) is a trajectory of ${\cal V}$, parametrized by $x$, which satisfies

\eq {\bf X}' = {\bf F} ({\bf X}) = \left( 
\begin{array}{c} 
h'\\
h''\\
h'+f(h)\\
\end{array}
\right) . \label{spaceq} \qe

\begin{figure}
\begin{center}
\psfrag{h}{$h$}
\psfrag{hp}{$h'$}
\psfrag{hpp}{$h''$}
\psfrag{legA}{$\text{Ca}^*=1.2$}
\psfrag{legB}{$\text{Ca}^*=0.1$}
\psfrag{legC}{$\text{Ca}^*=0.8$}
\psfrag{legD}{$\text{Ca}^*=1.0245$}
\includegraphics[width=12cm]{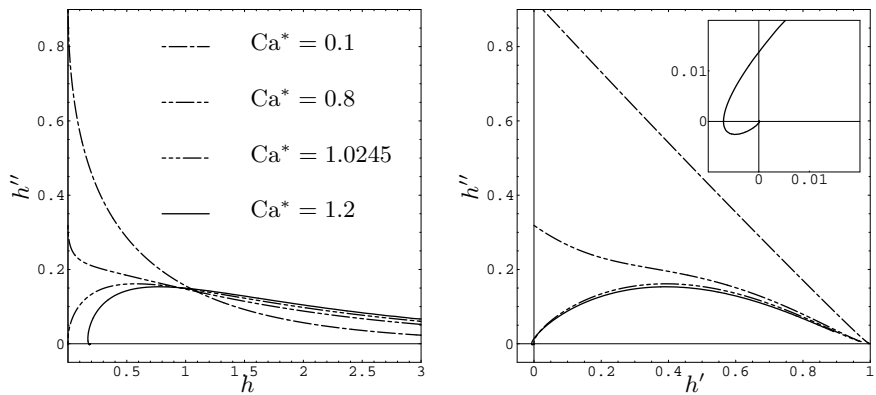}
\caption{Numerical solutions of Equation (\ref{totalrs}) for various capillary numbers, represented in the phase 
space (projected on the $(h,h'')$- and $(h',h'')$-planes). The rescaled tilt angle $\theta^*$ is fixed to $1$. 
The trajectories correspond to the physical solutions shown in Figure \ref{contlinesol}: for subcritical 
capillary numbers (the three dashed curves), the trajectory starts at a point on the $h''$-axis which 
corresponds to the contact line. On the opposite, the solid line correspond to a film solution, and thus 
does not cross the $h''$-axis. The insert shows the projection of the film solution in the $(h',h'')$-plane, at 
smaller scale.}
\label{TraPhaSpa}
\end{center}
\end{figure}

Some trajectories (the same as on Figure \ref{contlinesol}) are represented on Figure \ref{TraPhaSpa}. Notice that the film solution (solid curve) winds exponentially around a \emph{fixed point} on the $h$-axis.

\subsubsection{Hydrostatic solutions in the phase space}\label{speshootmet}

For large $x$, following the reasoning of Section \ref{hydrosol}, Equation (\ref{spaceq}) becomes linear:

\eq {\bf X}' = 
\left( 
\begin{array}{c c c} 
0 & 1 & 0\\
0 & 0 & 1\\
0 & 1 & 0\\
\end{array}
\right) {\bf X}
-
\left( 
\begin{array}{c} 
0\\
0\\
\theta^*\\
\end{array}
\right)
\label{spaceqmeniscus} .\qe

Any solution of (\ref{spaceqmeniscus}) which satisfies the boundary condition (\ref{BCip}) is included in a 
plane called $E_{\infty}$, which may be parameterized by $x$ and the apparent contact angle 
$\theta^*_{\text{ap}}$ introduced in Section \ref{hydrosol}. $E_{\infty}$ is defined by the equation 
$h'-h''=\theta^*$.

The solutions of the full Equation (\ref{spaceq}) which satisfy (\ref{BCip}) are included in a 
two-dimensional manifold, called $W$. This manifold tends to $E_{\infty}$ for large $h$. This allows us to 
approximate numerically the LLD film trajectories, for which we impose boundary conditions at 
$x \rightarrow -\infty$ and $x \rightarrow +\infty$. We may indeed use a shooting method with initial 
conditions varying along a constant (large) $h$ line on $E_{\infty}$. The boundary condition at 
$x \rightarrow +\infty$ is then approximatively satisfied at any step. The shooting method provides an 
approximation of the only solution that remains constant as $x$ tends to $-\infty$.

\begin{figure}
\begin{center}
\psfrag{h}{$h$}
\psfrag{hpp}{$h''$}
\includegraphics[width=10cm]{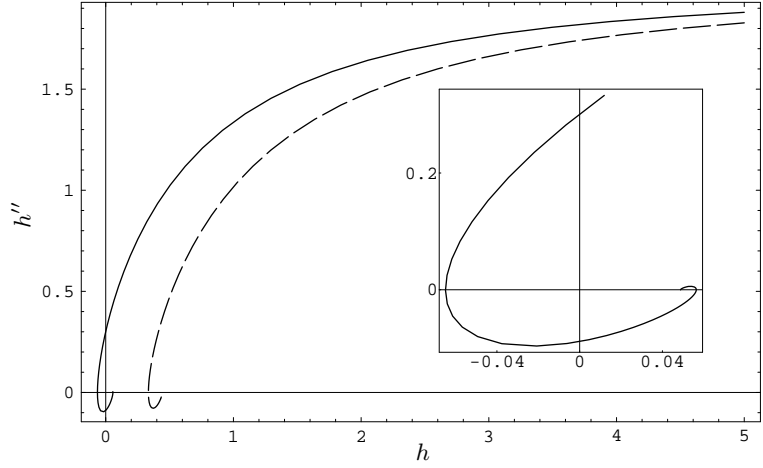}
\caption{Intersection of $W$ (the set of trajectories that tend to a horisontal water surface as $x$ tends to $+\infty$) with the $(h,h'')$ plane. These curves were obtained by a shooting method with $\theta^*=2$. Solid line: $\text{Ca}^*=2.1$; dashed line: $\text{Ca}^*=3$. Insert shows the solid curve at smaller scale.}
\label{interW}
\end{center}
\end{figure}

Figure \ref{interW} represents the intersection of $W$ with the $(h,h'')$ plane defined by $h'=0$, obtained 
by the shooting method, for two different capillary numbers, above and below $\text{Ca}^*_c$, for 
$\theta^*=2$. We observe numerically that the major effect of an increase in $\text{Ca}^*$ is a translation 
in the higher $h$ direction. The disappearance of the contact line solution may be described in the following 
way: any contact line trajectory is embedded in $W$, and the boundary conditions impose that it starts on 
the $h''$-axis, consequently, it can exist only if there is an intersection between $W$ and the $h''$-axis. 
Since the main effect of an increase of $\text{Ca}^*$ on $W$ is a translation along the $h$-axis, this 
intersection disappears above some value $\text{Ca}^*_c$ of the capillary number. Thus, the existence of a 
fixed point creates a \emph{separatrix} on the boundary $W$, that would otherwise be defined over the whole 
$(h,h')$-plane, which allows for the sudden disappearance of its intersection with the $h''$-axis.

\subsubsection{Fixed points}\label{fixpt}

A fixed point ${\bf X}_f$ in phase space corresponds physicaly to a film of constant height $h_f$:

\[{\bf X}_f = \left( 
\begin{array}{c} 
0\\
0\\
h_f\\
\end{array}
\right) .\]

The existence and values of fixed points depend on the parameters $\theta^*$ and $\text{Ca}^*$, as presented in Table \ref{fixedpoints}. In the following, we will focus on the largest fixed point ${\bf X}_f^+$, since it is the only one that may be acceptable physically (that is $h_f>0$). Let us linearize Equation (\ref{spaceq}) around ${\bf X}_f^+$:

\[{\bf X}' ={\bf J}_f^+\left( {\bf X} - {\bf X}_f^+ \right) \label{fpeq},\]

where ${\bf J}_f^+$ is the jacobian of ${\bf F}$ evaluated at ${\bf X}_f^+$, that is

\[{\bf J}_f^+ = \left( 
\begin{array}{c c c} 
0 & 1 & 0\\
0 & 0 & 1\\
f'(h_f^+) & 1 & 0\\
\end{array}
\right) .\]

\begin{table}
\begin{center}
\begin{tabular}{|c|c|}
\hline
Condition & Fixed points $h_f$ \\
\hline
$\text{Ca}^*<\frac{\theta^*}{4}$ & $\emptyset$ \\
\hline
$\text{Ca}^*=\frac{\theta^*}{4}$ & $-\frac{3}{2}$ \\
\hline
$\text{Ca}^*>\frac{\theta^*}{4}$ & $\frac{3}{2} \left( -1 \pm \frac{1}{\sqrt{3}} \sqrt{\frac{4
\text{Ca}^*}{\theta^*}-1}\right)$ \\
\hline
\end{tabular}
\caption{Existence and values of the fixed points of equation (\ref{spaceq}).}
\label{fixedpoints}
\end{center}
\end{table}

The local behaviour of solutions around the fixed point depends on the eigenvalues of ${\bf J}_f^+$, which 
are presented in Table
\ref{eigenvalues}. If the eigenvalues are real numbers, one is negative and the two others positive. 
So there is an unstable manifold of dimension two where the trajectories tend monotonically to the fixed 
point as $x$ tends to $-\infty$ and a stable manifold $S$ (separatrix) of dimension one. On the other hand, 
when the two eigenvalues are complex conjugate their common real part is always positive, and the 
trajectories in the corresponding unstable manifold wind around the fixed point while diverging from it 
at exponential rate. In physical space, the fluid surface forms damped stationary waves along the plate 
(see Figure \ref{ContactAtMinimum}). Therefore, the behaviour of $W$ in the vicinity of the fixed point may 
also be described by this linear expansion. Depending on the parameters, $W$ may either be defined over the 
whole $(h,h')$-plane, or tend to the separatrix $S$ (which ends on the fixed point). In the latter case, $W$ 
winds around the separatrix (as shown in Figure \ref{interW}) or tends to it monotonically. These various regimes are represented in the diagram of Figure \ref{phasedia}.

\begin{table}
\begin{center}
\begin{tabular}{|c|c|}
\hline
Condition & Eigenvalues \\
\hline
$\text{Ca}^* \in \Upsilon (\theta^*)$ & $a$, $b$, $c$ \\
\hline
$\text{Ca}^* \in \partial \Upsilon (\theta^*)$ & $a$, $-a/2$, $-a/2$\\
\hline
$\text{Ca}^* \not \in \Upsilon (\theta^*)$ & $a$, $-a/2 + i \Omega$, $-a/2 - i \Omega$ \\
\hline
\end{tabular}
\caption{Eigenvalues of the jacobian ${\bf J}_f^+$ at the largest fixed point. $a$, $b$, $c$ and $\Omega$ are real numbers and $\Upsilon (\theta^*) = \left[ \frac{9
\theta^{*3}}{2} \left( 1-\sqrt{1-\frac{1}{9\theta^{*2}}} \right) \; ;
\; \frac{9 \theta^{*3}}{2} \left( 1+\sqrt{1-\frac{1}{9\theta^{*2}}} \right) \right]$. When the eigenvalues are real, they satisfy: $a\;<\;0\;<\;b\;<\;c$.}
\label{eigenvalues}
\end{center}
\end{table}

\subsubsection{Critical capillary number}\label{CritCapNum}

\begin{figure}
\begin{center}
\psfrag{Ca}{$\text{Ca}^*$}
\psfrag{Cacr}{$\text{Ca}^*_c$}
\psfrag{theta}{$\theta^*$}
\psfrag{00}{$\emptyset$}
\psfrag{pc}{$(a+)$}
\psfrag{mc}{$(a-)$}
\psfrag{pr}{$(b+)$}
\psfrag{mr}{$(b-)$}
\psfrag{P}{$P$}
\includegraphics[width=10cm]{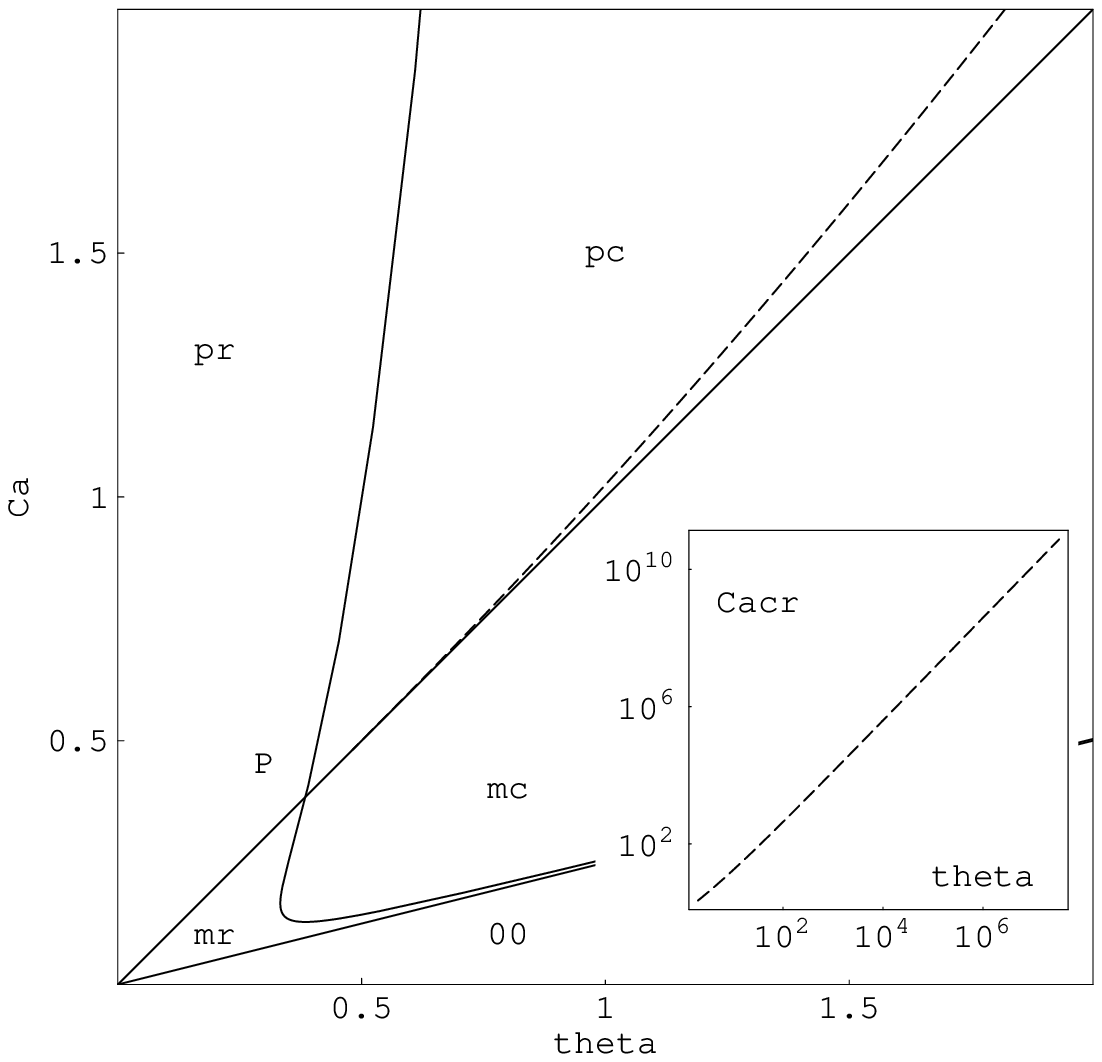}
\caption{Behaviour of $W$ close to the largest fixed point ${\bf J}_f^+$. $\emptyset$: no fixed
point; $+$: positive fixed point; $-$: negative fixed point; $a$: spiraling trajectories;
$b$: monotonic trajectories. Continuous lines delimit the various behaviors. The dashed line represents the critical capillary number above
which the contact line solution disappear $\text{Ca}^*_c$. Insert: asymptotic behavior of the critical capillary number for large $\theta^*$, (logarithmic scale).}
\label{phasedia}\label{CacrLog}
\end{center}
\end{figure}

At any point on the $h$-axis, ${\bf X}'$ is parallel to the $h''$-axis (see equation (\ref{spaceq})). Consequently, the intersection of $W$ with the $(h,h'')$ plane (represented on Figure \ref{interW}) has vertical tangent vectors whenever it crosses the $h$-axis. \label{proofcnull} This explains the behaviour of $h''(0)$ close to the critical capillary number (see Figure \ref{hopf}), which may be interpreted as a saddle-node bifurcation. This property is useful for the numerical determination of $\text{Ca}^*_c$ at a given $\theta^*$: since we know that the second derivative $h''(0)$ must vanish at the critical capillary number, we may approximate $\text{Ca}^*_c$ by a shooting method which varies $\text{Ca}^*$ for constant initial conditions (that is, $h=h'=h''=0$). We show in Figure \ref{phasedia} the evolution of the critical capilary number with $\theta^*$, together with the diagram showing the different regimes described above. Notice that since $\theta^*$ is a rescaled parameter, we have been able to investigate a large range of values, up to $\theta^* \approx 10^7$.

\begin{figure}
\begin{center}
\psfrag{h}{$h$}
\psfrag{x}{$x$}
\psfrag{A}{$A$}
\psfrag{B}{$B$}
\psfrag{C}{$C$}
\psfrag{D}{$D$}
\psfrag{c}{$h''$}
\psfrag{spac}{} 
\psfrag{Ca1}{$\text{Ca}^* \approx 241.3$}
\psfrag{Ca2}{$\text{Ca}^* \approx 195.5$}
\psfrag{Ca}{$\text{Ca}^*$}
\psfrag{Cacr}{$\text{Ca}^*_c$}
\includegraphics[width=10cm]{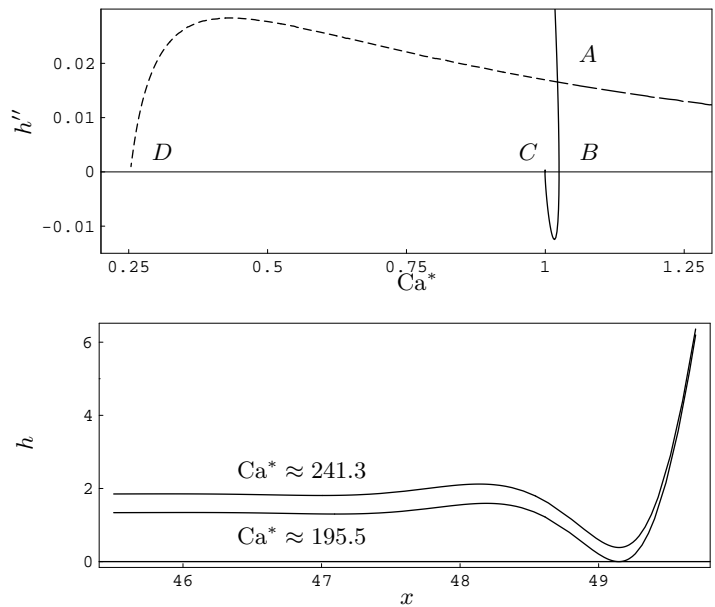}
\caption{ \emph{Above:} Second derivative of the film height against the capillary number, for $\theta^* =1$. The solid line corresponds to $h''(0)$ for the contact line solution. The dashed line represents the second derivative of $h$ at the point $x_{\text{min}}$ where in the case of a LLD film solution, the film is thinnest. Large dashes are used if $h(x_{\text{min}}) > 0$, and short dashes otherwise.\newline
Points $A$,$B$,$C$ and $D$ correspond respectively to the following values of  $\text{Ca}^*$: $\text{Ca}^*_{c,2}$, $\text{Ca}^*_{c,1}$, $\theta^*$ and $\theta^*/4$. \newline
\emph{Below:} Two LLD film solutions for $\theta^* \approx 66.62$ at different capillary numbers. The curve below correspond to point $A$, since $h(x_{\text{min}})$ becomes negative.}
\label{ContactAtMinimum}
\end{center}
\end{figure}

We did not find any reason for the disappearance of the contact line solution to coincide with the appearance of the LLD film solution as $\text{Ca}^*$ is varied. It may well be possible that, as $W$ has already intersected the $h''$-axis, the LLD film trajectory rolls up around the fixed point without $h$ ever becoming negative. Some numerical simulations give us confidence that hysteresis indeed occurs (that is $\text{Ca}^*_c>\text{Ca}^*_{c,2}$): a slight hysteresis may indeed be observed in Figure \ref{ContactAtMinimum}. Again, the present study  is limited to LLD films of null flux, and other solutions may exist for the same parameters values. Thus, the hysteresis here observed can only describe a reduced part of the solutions set.

When the capillary number is decreased from a supercritical value, the height of the stationary film $h_f$ decreases, and eventually the film thickness vanishes at some point $x_{min}$ (see Figure \ref{ContactAtMinimum}). This point must be a minimum and in that case both $h(x_{min})$ and $h'(x_{min})$ vanish, so this film solution is also a contact line solution. This explains the change of the sign of $h(x_{\text{min}})$ observed at point $A$ in Figure \ref{ContactAtMinimum}.

The asymptotic behavior of $\text{Ca}^*_c$ at large $\theta^*$ has also been investigated (Figure \ref{CacrLog}). We find that the critical capillary number behaves asymptotically as a power law of the tilt angle, that fits to:

\eq \text{Ca}^*_c \mathop{\sim}_{\theta^* \rightarrow \infty} 0.3936 \: \theta^{*\;1,4998} . \label{powerlaw} \qe

This suggest that for high $\theta^*$, $\text{Ca}^*_c$ behaves like $\theta^{*\;3/2}$.

\section{Asymptotic results}
\label{analytical}
In the following section we describe a rough analytical approach, inspired from that of \cite{egg04}, which leads to the power law (\ref{powerlaw}) for the capillary number obtained numerically in the previous section.

\subsection{Overview}

To determine the dependence of the critical capillary number with the angle, we need to better
understand how the solution near the contact line connects with the free surface at infinity. 
We therefore seek to determine the matching between these two domains.
This has actually been done when considering the classical plate withdrawal problem. It involves a matching between three
regions: one near the contact line, a capillary-viscous one and the gravity-capillary interface \cite[]{egg04}.
However, two major differences arise in the present case compared to the usual problem. First, equation (\ref{finalset})
is regular over the whole range $h \in [0 ; + \infty[$, whereas in the usual problem Navier slip condition (\ref{slip}) leads to a
pressure divergence at the contact line. However, it has been shown (in the case of an \emph{advancing} contact line) that the exact form of the slip law near the contact 
line does not influence the matching procedure with the far-field solution of the free surface \cite[]{egg04b,duss74}. 
The second difference lies in the contact angle condition that we consider null instead of small but finite for solid 
plates. Such a condition is crucial as it can be seen from \cite[]{egg04} where the solutions are expanded in powers 
of the small parameter $\text{Ca} / \theta_e^3 $ ($\theta_e $ being the static contact angle). Consequently, we cannot
obtain a proper matching between the behavior in the contact line zone (cubic polynomial at leading order) with the 
famous logarithmic behavior in the capillary-viscous region:

\[ h'(x)=[9 \text{Ca} \ln (\frac{\pi}{2^{2/3} \beta^2 x})]^{1/3} \]

However, we can bypass this difficulty by a slight change in the equations leading to a single approximation valid over
the first two regions (contact line zone and capillary-viscous one).

\subsection{A two-zones matching}

The procedure hereafter presented is based on the assumption that the linear term in the right-hand term 
denominator of equation (\ref{finalset}) is not of fundamental importance. In particular, the coefficient 
$\alpha$ was arbitrarily set to one at the beginning of this study, but numerical investigations have shown 
that it may be set to much different values, changing the results of only a few percent. For values of 
$1 / \alpha$ greater than $2/ \sqrt{3}$, equation (\ref{finalset}) becomes singular for \emph{negative} 
values of $h$, which are physically meaningless. We therefore set from now on $1/\alpha = 2/ \sqrt{3}$ without 
any change in the equation properties, so that equation (\ref{finalset}) reduces to

\eq h''' - h' + \theta^* = \frac{\text{Ca}^* }{(h / \sqrt{3}+1)^2}. \label{eqred} \qe

Close enough to the contact line, the film slope $h'$ may be neglected due to the boundary condition (\ref{finalset}). In addition, close to the critical capillary number, $\text{Ca}^* \gg \theta^*$ (this is suggested by the assymptotic behavior (\ref{powerlaw})), and equation (\ref{eqred}) becomes

\[ h'''= \frac{\text{Ca}^* }{(h / \sqrt{3}+1)^2}, \]

solved analytically (as performed by \cite{duffy}) after the rescaling 

\[ x= \frac{\sqrt{3}}{(3 \text{Ca}^*)^{1/3}} \xi, \; h(x)= \sqrt{3} (y(\xi)-1). \]

This rescaling leads to Tanner's problem:

\[ y'''=\frac{1}{y^2}, \; y(0)=1, \; y'(0)=0.\]

Its solution may be parametrized in terms of Airy functions $\text{Ai}$ and $\text{Bi}$:

\eq
\xi=2^{1/3} \frac{\text{Bi}(s_0) \text{Ai}(s) - \text{Bi}(s) \text{Ai}(s_0)}{\text{Bi}'(s_0) \text{Ai}(s) - \text{Bi}(s) \text{Ai}'(s_0)}, \; 
y_{\text{in}}=\frac{1}{\pi^2 (\text{Bi}'(s_0) \text{Ai}(s) - \text{Bi}(s) \text{Ai}'(s_0))^2},
\label{airysol}
\qe

where $s_0$ is an integration constant, and $s$ varies between consecutive solutions of equation

\eq \text{Bi}'(s_0) \text{Ai}(s) - \text{Bi}(s) \text{Ai}'(s_0)=0 .\label{sdefine}\qe

It was shown in Section \ref{proofcnull} that, at the critical capillary number, $h''$ vanishes (and so does $y''$). This property sets $s_0$ to zero and the range for $s$ to $[s_1,0[$, where $s_1 \approx -1.98635$ is the largest solution to equation (\ref{sdefine}). Matching solution (\ref{airysol}) with the meniscus solution should provide a condition on $\text{Ca}^*$ and $\theta^*$. At large $\xi$, the behavior of $y_{\text{in}}$ is

\[ y_{\text{in}} = a \xi^2 + b \xi + c + O(\frac{1}{\xi}) \]

where $a$,$b$ and $c$ have the following expressions (if we define $z(s)= \pi (\text{Bi}(0) \text{Ai}(s) - \text{Bi}(s) \text{Ai}(0))$) :

\[ a = \left( \frac{\text{Bi}'(0)}{2^{1/3} \text{Bi}(s_1)} \right)^2 \approx 0.758947, \;
b = 2a \left( \frac{2^{1/3}\text{Bi}'(s_1)}{\text{Bi}'(0) z'(s_1)} \right) \approx 1.12697, \]
\[ c = \frac{b^2}{4a} + \frac{\text{Bi}'(s_1) z''(s_1) - \text{Bi}''(s_1) z'(s_1)}{\text{Bi}(s_1) z'^3(s_1)} \approx 2.06713.\]

Matching $y_{\text{in}}$ with the second-order Taylor expansion of $h_{\infty}$ for vanishing $x$ leads to:

\[ \theta^*-\theta^*_{\text{ap}}= 2a 3^{1/6} \text{Ca}^{* \; 2/3}_c, \;  A_{\infty}=\sqrt{3}(c-1)-\theta^*+\theta^*_{\text{ap}}, \;
\theta^*_{\text{ap}}= b (3 \text{Ca}^*_c)^{1/3}. \label{matchedeq} \]

When the critical capillary number tends to infinity, the first of the previous equations reads

\eq \text{Ca}^*_c \mathop{\sim}_{\theta^* \rightarrow \infty} \frac{1}{(2a)^{3/2} 3^{1/4}} \: \theta^{* \; 3/2}. \label{powerlawth} \qe

\begin{figure}
\begin{center}
\psfrag{x}{$x$}
\psfrag{h}{$h$}
\includegraphics[width=10cm]{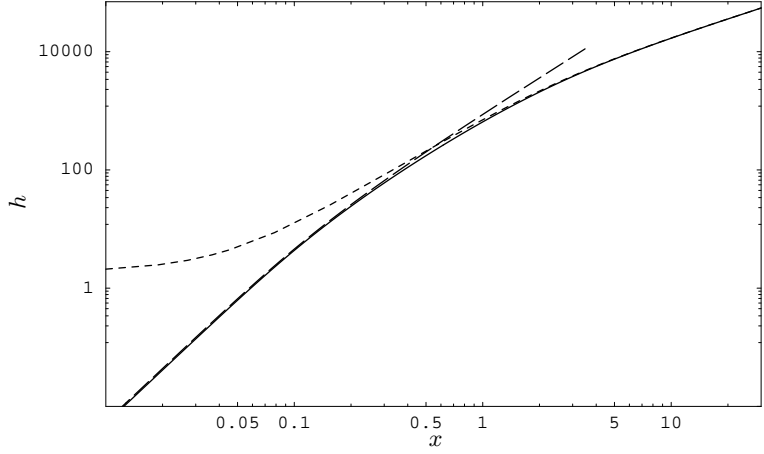}
\caption{Comparison between numerical results (solid line) and the matching presented in this paper (long and short dashes), for $\theta^* \approx 1886.19$ and $\text{Ca}^* \approx  \text{Ca}^*_c \approx32871.1$, in logarithmic scales. Long dahes: analytical solution to Tanner's problem; short dashes: capillary-gravity meniscus.}
\label{MatchingTest}
\end{center}
\end{figure}

This fits remarkably well with the numerical estimation (\ref{powerlaw}), since $ 1/((2a)^{3/2} 3^{1/4}) \approx 0.40631 $ (the above numerical fit giving $0.3936$). The matching is compared to numerical results in Figure \ref{MatchingTest} for a high value of $\theta^*$, and we observe a remarkable agreement with a reasonably large overlap region.

Moreover, this matching procedure provides the following law for the apparent contact angle, at the critical speed: $ \theta_{\text{ap}}= b (3 \text{Ca}_c)^{1/3} $ (notice that the rescaling term $\lambda$ disappears). This is reminiscent of the famous Tanner law which relates the contact line velocity to the apparent contact angle \cite[]{tanner}.

\section{Discussion and conclusions}
\label{conclusion}
In this work, a continuum model of the forced dewetting on a porous material has been presented. In the framework of lubrication, an ordinary non-linear differential equation was derived, close to the one investigated by \cite{hock01}. Even if the (microsopic) contact angle is assumed to vanish, a stationary contact line is found to exist for low dewetting velocity. Moreover, a transition between this steady contact line and the deposit of a LLD-film must occur, since there is a critical capillary number above which no contact line solution can exist.

In the present study, the apparent contact angle (as defined by \cite{egg04}, that is $\theta^*_{\text{ap}} \lambda$ in the present notations), does \emph{not} vanish at the critical capillary number (see Section \ref{analytical}). This behavior is different from the one obtained by \cite{egg04}, where the apparent contact angle was found to vanish at the transition from a contact line to a LLD film in the case of finite microscopic contact angle $\theta_e$. This discrepancy is surprising since one would expect the problem studied by \cite{egg04} to correspond with the present one when $\theta_e \rightarrow 0$ and $\lambda \rightarrow 0$. However, in the latter limit the rescaling by $\lambda$ used here cannot hold. Also, as previously mentioned, the expansion in powers of $\text{Ca}/\theta_e^3$ performed by \cite{egg04} becomes ill-defined.

\begin{figure}
\begin{center}
\psfrag{S1}{$ S_c=0.05 $}
\psfrag{S2}{$ S_c=0.12 $}
\psfrag{k1}{$ k=10^{-12} \; \text{m}^2 $}
\psfrag{k2}{$ k=900\:.\:10^{-12} \; \text{m}^2 $}
\psfrag{theta}{Tilt angle [degrees]}
\psfrag{V}{\bigskip Plate Velocity [$\text{cm.s}^{-1}$]}
\includegraphics[width=10cm]{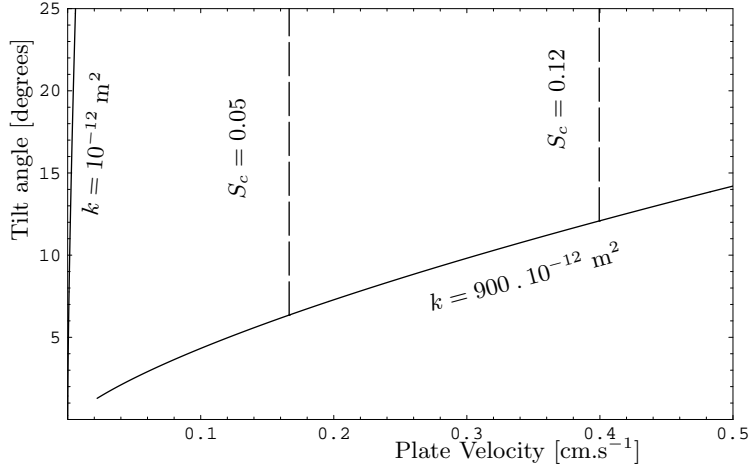}
\caption{Critical speed for two values of the permability $k$. For each value of the permeabiblity, a contact line can exist only \emph{above} the corresponding solid curve. The dashed line represents constant Shields numbers $S_c$. The values of the physical parameters used in this model are those of Daer \emph{et al.} \cite[]{dae03}: $g=9.81 \; \text{m.s}^{-2}$, $\rho=1000 \; \text{kg.m}^{-3}$,  $\rho_g=2750 \; \text{kg.m}^{-3}$, $\eta=10^{-3} \; \text{kg.m}^{-1}.\text{s}^{-1}$ and $\gamma_s=0.07 \; \text{N.m}^{-1}$. The velocity and inclination ranges are those of the experiment.}
\label{DiagrDaer}
\end{center}
\end{figure}

Different conclusions can be drawn from our results regarding the erosion experiment performed by \cite{dae03}. First, the existence diagram of the contact line can be drawn using the experimental values of the physical parameters. In the present theory, the permeability $k$ of the porous material is crucial, as is the characteristic slip length at the solid-liquid interface. The value of this parameter may depend strongly on the compaction of the granular material (say between $10^{-12}\; \text{m}^2$ and $900\: . \:10^{-12}\; \text{m}^2$, respectively the value measured by \cite{dae03} and the square of the grain size). Figure \ref{DiagrDaer} presents the critical velocities obtained for these two extremal values of the permeability. For the lowest permeability, and down to the smallest withdrawal velocities of the erodible plate, no contact line can exist. On the other hand, when choosing the largest permeability, the critical speed line is of order of those of the experiement.

The flow acts on the granular medium mainly through the bottom shear rate $\tau=\partial u/\partial y$, which is known to trigger the erosion process (see \cite{charru}). From this shear rate, we can define the Shields number $S$, which compares the viscous force applied to the grains by the flow, to gravity force:

\[ S=\frac{\eta \tau }{(\rho_g - \rho) g d}, \]

where $\rho_g$ is the density of the grains (this expression stands only for small angles). Though the erosion prossess on a granular bed results from discrete and complex phenomena, classical models assume that it starts at a threshold value of the shear rate, at which a critical Shields number $S_c$ is defined \cite[]{charru}. A typical value for $S_c$ is $0.05$ (see among others \cite{fredsoe}), but \cite{dae03} used $S_c=0.12$ to fit their data. In addition to this large range of possible values, note that the critical Shields number is a function of the slope of the bottom: the more inclined it is, the easier it is for the flow to lift grains, thus the tilt reduces the value of $S_c$. The shear rate may be deduced from our model, as a function of the dimensionless height of the film $h$:

\[ \tau=\frac{U h}{\sqrt{k}(h^2/3 + h + 1)}. \]

This expression admits a maximum value $\tau_{\text{max}}=U/(\sqrt{k}(1+2/\sqrt{3}))$ for $h=\sqrt{3}$, that is necessarily reached in the case of a contact line, since $h$ stretches from zero to infinity. If the Shields number is assumed to be independent of the tilt angle of the plate (to first order), $S=S_c$ defines a vertical line in Figure \ref{DiagrDaer} (represented only for $ k=900\:.\:10^{-12} \; \text{m}^2 $). On the left of such a line, no erosion should occur since the Shields number is smaller than the critical value, thus to account for the erosion patterns observed by \cite{dae03} at small velocities, a low value of $S_c$ is required.

Now, if a LLD film covers the plate, $h$ admit a minimum value $h_{min}$ which is found numerically to increase with the plate velocity. Once $h_{min}$ is larger than $h=\sqrt{3}$, the maximum value of the shear rate becomes smaller than $\tau_{\text{max}}$, and could decrease as the capillary number increases. Thus, the transition from contact line to LLD film could induce a strong change in the stress regime. If the physical parameters are choosen in the experimental range, no stress jump is observed numerically at the transition, but two assumptions should be relaxed in order to evaluate precisely the bottom stress close to the critical capillary number: the permanent regime, and the constraint of null outflow ($Q=0$). In particular, if a negative outflow appears in reality (that is, if water is withdrawn from the bath), the shear rate should be reduced. Such a sharp stress variation could provide an explanation for the transition between the different erosion patterns observed by \cite{dae03}.

At the transition from contact line to LLD film, transient regimes should not be ignored. They have been studied in the litterature for non-vanishing contact angles \cite[]{hock01}, and future studies will aim at understanding the case of null contact angle, which has been shown here to be quite different.

It is our pleasure to thank Daniel Lhuillier, Pierre-Yves Lagrée, Eric Clément, Florent Malloggi and Jens Eggers for stimulating discussion.

\appendix

\section{Pressure divergence at the contact line}
\label{DivergenceDemo}
We aim here to demonstrate briefly that the classical Navier slip condition, leading to Equation (\ref{slip}), is not sufficient to eliminate all the singularities at the contact-line, even for a non-vanishing microscopic contact angle ($\theta_e > 0$). Indeed, the first order expansion of Equation (\ref{slip}) is \cite[]{egg04,hock83}

\[ h'(x) \sim \theta_e - \frac{3 \text{Ca}}{\theta_e^2} \left( 1 + \ln  \left( \frac{x \theta_e}{\lambda_N} \right) \right), \]

where $h$ is dimensionless but not rescaled (that is, h is the height of the water surface divied by the capillary length). In the lubrication approximation that we have used throughout, the pressure at the plate reads

\[ \left. p \right|_{y=0} = \rho g \cos (\theta) h - \gamma h'', \]

and since $ h''(x) \sim - 3 \text{Ca} / (\theta_e^2 x) $ the pressure diverges at the contact line. In the present paper, due to the permeability of the porous plate, the expansion of $h$ near the contact-line is a third order polynomial ($h \sim h''(0) x^2/2 + (\text{Ca}^*-\theta^*)x^3/6$), and thus the pressure does not diverge.

\section{Derivation of the fundamental equation}
\label{MoveEq}
In the following, we aim to derive Equation (\ref{total}) from the two-dimensional Navier-Stokes equation. $x$ and $y$ refer to the axes of Figure \ref{schema}. In the frame of the lubrication approximation, and assuming both a permanent regime and small Reynolds number, momentum conservation reads

\eq  - \frac{1}{\rho} \frac{\partial p}{\partial x} + g \sin (\theta) + \nu \frac{\partial^2 u}{\partial y ^2} =0 \label{hqdm} \qe

\[  - \frac{1}{\rho} \frac{\partial p}{\partial y} - g \cos (\theta) =0, \]

where $u$ and $v$ stand for the water velocity components respectively parallel and perpendicular to the plate. The second equation may be integrated to give

\[ p = \rho g \cos (\theta) (h-y) + p_L \]

where $p_L$ is the pressure due to surface tension. Now, if $\theta$ is small enough, the slope $h'$ of the free surface should remain reasonably small, so that $h''$ approximates its curvature, and $p_L \approx - \gamma h''$ (and, similarly, $ \sin (\theta) \approx \theta$ and $\cos (\theta) \approx 1$ at first order). The boundary conditions on $u$ at the bottom and the top of the film are (see Equations (\ref{nostress}), (\ref{slipporous}) and (\ref{darcy}))

\[ \left. u  \right| _{y=0} + U = \frac{\sqrt{k}}{\alpha} \left. \frac{\partial u}{\partial y} \right| _{y=0} - \frac{k}{\eta} \left( \left. \frac{\partial p}{\partial x}  \right| _{y=0} - \rho g \theta \right) \]

\[ \left. \frac{\partial u}{\partial y} \right| _{y=h(x)} = 0, \]

thus integrating Equation (\ref{hqdm}) we obtain

\eq u= \frac{1}{\nu} \left( \frac{1}{\rho} \frac{\partial p}{\partial x} - g \theta \right) \left( \frac{y^2}{2} - h y 
- \frac{\sqrt{k}}{\alpha} h - k \right) + U.
\label{ufield} \qe

The mass flux is

\[ \rho \int_0^h  u \; dy = Q. \]

In steady state, mass conservation imposes that $Q$ is a constant which vanishes in the case of a contact line. This condition provides the non-linear equation studied throughout this paper: $Q=0$ reads

\[\frac{\gamma}{\rho} h''' - g h ' + g \theta = \frac{\nu U}{ h^2/3 + h \sqrt{k}/ \alpha + k.} \]

which, after rescaling $x$ and $h$ by the capillary length $l_c$, reduces to Equation (\ref{total}).

\bibliographystyle{jfm}
\bibliography{biblioLLDJFM}

\end{document}